\documentclass[12pt]{article}
\begin{document}

\centerline{\Large\bf  Null Surfaces in Static Space-times}
\vskip .7in
\centerline{Dan N. Vollick}
\vskip .2in
\centerline{Irving K. Barber School of Arts and Sciences}
\centerline{University of British Columbia Okanagan}
\centerline{3333 University Way}
\centerline{Kelowna, B.C.}
\centerline{Canada}
\centerline{V1V 1V7}
\vskip 0.5in
\centerline{\bf\large Abstract}
\vskip 0.5in
\noindent
In this paper I consider surfaces in a space-time with a Killing vector $\xi^{\alpha}$ that is time-like and hypersurface orthogonal on one side of the surface. The Killing vector may be either time-like or space-like on the other side of the surface. It has been argued that the surface is null if $\xi_{\alpha}\xi^{\alpha}\rightarrow 0$ as the surface is approached from the static region. This implies that, in a coordinate
system adapted to $\xi$, surfaces with $g_{tt}=0$ are null. In spherically symmetric space-times the
condition $g^{rr}=0$ instead of $g_{tt}=0$ is sometimes used to locate null surfaces.

In this paper I examine the arguments that lead to these two different criteria and show that both arguments are incorrect.
A surface $\xi=$ constant has a normal vector whose norm is proportional to $\xi_{\alpha}\xi^{\alpha}$. This lead to the
conclusion that surfaces with $\xi_{\alpha}\xi^{\alpha}=0$ are null. However, the proportionality factor generally diverges
when $g_{tt}=0$, leading to a different condition for the norm to be null. In static spherically symmetric
space-times this condition gives $g^{rr}=0$, not $g_{tt}=0$.

The problem with the condition $g^{rr}=0$ is that the coordinate system is singular on the surface.
One can either use a nonsingular coordinate system or examine the induced metric on the surface to determine if it is null.
By using these approaches it is shown that the correct criteria is $g_{tt}=0$. I also examine the condition required for
the surface to be nonsingular.

\newpage

\section{Introduction}
In this paper I consider surfaces in a space-time with a Killing vector $\xi^{\alpha}$ that is time-like and hypersurface orthogonal on one side of the surface (the space-time is therefore static in this region). The Killing vector may be either time-like or space-like on the other side of the surface.
It has been argued \cite{Vi1} that the surface is null if
\begin{equation}
\xi_{\alpha}\xi^{\alpha}\rightarrow 0
\end{equation}
as the surface is approached from this static region (for simplicity I will often state this condition as $\xi_{\alpha}\xi^{\alpha}=0$).
In a coordinate system in which the metric is given by
\begin{equation}
ds^2=g_{tt}(\vec{x})dt^2+g_{ij}(\vec{x})dx^idx^j
\label{metric}
\end{equation}
this condition corresponds to $g_{tt}\rightarrow0$ as the surface is approached.

Now consider surfaces of constant $r$ in a spherically symmetric space-time with a metric
\begin{equation}
ds^2=g_{tt}(r)dt^2+g_{rr}(r)dr^2+r^2d\Omega^2\;.
\label{spherical}
\end{equation}
The normal to the surface is given by $n_{\mu}=(0,1,0,0)$ and its norm is $n_{\mu}n^{\mu}=g^{rr}$.
This seems to indicate that a null surface corresponds to $g^{rr}=0$, not to $g_{tt}=0$.
Both $g_{tt}=0$ \cite{Vi1,Th1} and $g^{rr}=0$ \cite{Ca1,Ta1} have been used to locate null surfaces.

In this paper I examine this issue in detail. I show that there is an problem with the argument that leads
to the conclusion that surfaces defined by $\xi_{\alpha}\xi^{\alpha}=0$
are null. This argument actually leads to the condition
\begin{equation}
g^{ij}g_{tt,i}g_{tt,j}=0
\end{equation}
for a surface of constant $\xi_{\alpha}\xi^{\alpha}$ to have a normal that is null.
In a spherically symmetric space-time, with $\partial_rg_{tt}\neq 0$, this condition is $g^{rr}=0$, which is consistent
with the result obtained below equation (\ref{spherical}).
However, it turns out that these approaches are incorrect. If $g^{rr}\rightarrow 0$ as a surface of constant
$r$ is approached the coordinate system is singular on the surface. One can
either use a nonsingular coordinate system or examine the induced metric on the surface to determine if it is null. By using these
approaches it will be shown that the correct criteria is $g_{tt}=0$. Since surfaces of infinite redshift also correspond to $g_{tt}=0$,
we see that any null surface in a static space-time must also be a surface of infinite redshift and vice-versa.

For a null surface to be nonsingular the Kretschmann scalar must be finite on the surface. It is shown that, in spherically symmetric space-times, the Kretschmann scalar
will diverge on a null surface unless $g^{rr}\rightarrow 0$ as the surface is approached. It is also shown that $g^{rr}$ can vanish and $g_{tt}$ can be finite and
nonzero on a nonsingular surface. Thus, $g^{rr}=0$ is a necessary but insufficient condition for a surface to be null and nonsingular.

One approach that has been used to find null surfaces in the Kerr space-time involves looking for surfaces defined by $f(r,\theta)$=constant
whose norm is null (see for example \cite{La1,Ad1}). The surfaces that are found using this approach are surfaces of constant $r$ with $g^{rr}=0$. However, the
coordinate system is singular on these surfaces, so this result needs to be checked. It is easy to verify that the induced metric
on these surfaces is null showing that they are null surfaces.

\section{Null Surfaces in Static Space-times}
In this section I will consider surfaces in a space-time with a Killing vector $\xi^{\alpha}$ that is time-like and hypersurface orthogonal on one side of the surface. The Killing vector may be either time-like or space-like on the other side of the surface.
One can always work in a coordinate system in which $\xi=\partial/\partial t$ and the metric takes the form given in
(\ref{metric}) in the static region.
It was argued in \cite{Vi1} that surfaces defined by $\xi^{\alpha}\xi_{\alpha}=0$ are null.
To examine this consider the normal to the surface $\xi_{\alpha}\xi^{\alpha}=$constant
\begin{equation}
n_{\mu}=\frac{1}{2}\nabla_{\mu}\left(\xi_{\alpha}\xi^{\alpha}\right)\;,
\label{normal}
\end{equation}
which we assume is nonvanishing ($\partial_kg_{tt}\neq 0$).
It was shown in \cite{Vi1} that
\begin{equation}
n_{\mu}n^{\mu}=\frac{1}{2}\left(\xi_{\alpha}\xi^{\alpha}\right)\left(\nabla_{\lambda}\xi_{\beta}\nabla^{\lambda}
\xi^{\beta}\right)
\end{equation}
and it was then concluded that null surfaces correspond surfaces with $\xi_{\alpha}\xi^{\alpha}=g_{tt}=0$.
However, the term $\nabla_{\lambda}\xi_{\beta}\nabla^{\lambda}\xi^{\beta}$ that multiplies
$\xi_{\alpha}\xi^{\alpha}$ is given by
\begin{equation}
\nabla_{\lambda}\xi_{\beta}\nabla^{\lambda}\xi^{\beta}=\frac{1}{2}g^{tt}g^{ij}(g_{tt,i})(g_{tt,j})\;.
\end{equation}
Thus,
\begin{equation}
n_{\mu}n^{\mu}=\frac{1}{4}g^{ij}g_{tt,i}g_{tt,j}\;,
\end{equation}
and does not generally vanish when $g_{tt}=0$. Note that this follows directly from (\ref{normal}).
The condition for a surface of constant $\xi_{\alpha}\xi^{\alpha}$ to have a null normal is, therefore, given by
\begin{equation}
g^{ij}g_{tt,i}g_{tt,j}=0\; .
\label{cond}
\end{equation}
In a spherically symmetric static space-time, with $\partial_rg_{tt}\neq 0$, this condition is $g^{rr}=0$, not $g_{tt}=0$.
This also follows directly from $n_{\mu}=(0,1,0,0)$ and $n_{\mu}n^{\mu}=g^{rr}$. However, we will see that it is incorrect to conclude that surfaces with
$g^{rr}=0$ are null.
If $g^{rr}\rightarrow 0$ as a surface of constant $r$ is approached the coordinate system is singular on this surface. One can
either use a nonsingular coordinate system or examine the induced metric on the surface to determine if it is null.
By using these
approaches it will be shown that the correct criteria in static spherically symmetric space-times is $g_{tt}=0$.
I also show that null surfaces correspond to $g_{tt}=0$ in a static space-time if the ``spatial" part of the
induced metric is nonsingular.

\section{Spherically Symmetric Static Space-times}
In this section I will examine null surfaces in a static spherically symmetric space-time with a metric given by (\ref{spherical}).
For simplicity I will assume that $g_{tt}g_{rr}\leq 0$ everywhere, so that there is no signature change in the space-time.
Consider the surface defined by $\Phi=r-r_0=0$. This surface has a normal $n_{\mu}=(0,1,0,0)$ and the norm of $n_{\mu}$ is given by
\begin{equation}
n_{\mu}n^{\mu}=g^{rr}\;.
\end{equation}
Thus, for the surface to be null it would appear that we require $g^{rr}=0$. However, this implies that the coordinate system is
singular on the surface.

Consider transforming to Eddington-Finkelstein coordinates. The null geodesics satisfy
\begin{equation}
t\pm\int\sqrt{-\frac{g_{rr}}{g_{tt}}}dr=constant
\end{equation}
and from this we introduce an ingoing null coordinate
\begin{equation}
v=t-\int\sqrt{-\frac{g_{rr}}{g_{tt}}}sign(g_{tt})dr\;.
\end{equation}
The $sign(g_{tt})$ is required so that we obtain the correct ingoing null coordinate and metric. For example in
the Schwarzschild case
\begin{equation}
-\sqrt{-\frac{g_{rr}}{g_{tt}}}sign(g_{tt})=\left|\left(1-\frac{2GM}{r}\right)^{-1}\right|sign\left(1-\frac{2GM}{r}\right)=\left(1-\frac{2GM}{r}\right)^{-1}\;.
\end{equation}
The Eddington-Finkelstein metric is given by
\begin{equation}
ds^2=g_{tt}dv^2+2\sqrt{-g_{tt}g_{rr}}dvdr+r^2d\Omega^2\;.
\label{14}
\end{equation}
Once again consider the normal $n_{\mu}=(0,1,0,0)$. Its norm is given by
\begin{equation}
n_{\mu}n^{\mu}=g^{rr}
\end{equation}
so that it still appears that we require $g^{rr}=0$. However, if $g^{rr}=0$ on the surface the coordinate system
is still singular unless $g_{tt}\rightarrow 0$ keeping $g_{tt}g_{rr}$ finite and nonzero.
In this case $g^{rr}\sim g_{tt}$ and either $g^{rr}=0$ or $g_{tt}=0$ can be used.

Consider the case when (\ref{14}) is singular. To obtain a nonsingular coordinate system define a new radial variable by
\begin{equation}
\bar{r}=\int\sqrt{-g_{tt}g_{rr}}dr\;.
\end{equation}
The metric in this new coordinate system is given by
\begin{equation}
ds^2=g_{tt}dv^2+2dvd\bar{r}+r(\bar{r})^2d\Omega^2\;.
\end{equation}
Now consider the normal $n_{\mu}=(0,1,0,0)$, which has norm
\begin{equation}
n_{\mu}n^{\mu}=-g_{tt}\;.
\end{equation}
Thus, in this nonsingular coordinate system the requirement for the surface to be null is $g_{tt}=0$. Since this result was
deduced in a nonsingular coordinate system the general requirement for a surface to be null
is $g_{tt}=0$ not $g^{rr}=0$. The reason for the discrepancy between the singular and nonsingular
coordinate systems is related to how the normal transforms. Consider the transformation from the
$(v,r,\theta,\phi)$ coordinates to the $(v,\bar{r},\theta,\phi)$ coordinates
\begin{equation}
\bar{n}_r=(-g_{tt}g_{rr})^{-1/2}n_r\;.
\end{equation}
Thus, a well defined normal in the singular coordinate system does not correspond to a well defined normal in the
nonsingular system. The transformation from the initial Schwarzschild coordinates to the $(v,r,\theta,\phi)$ Eddington-Finkelstein
coordinates does not change the normal, so the same requirement for a surface to be null is obtained in both of these coordinate systems.

Similar results can be found by transforming to Painlev\'{e}-Gullstrand coordinates. Consider a freely falling observer with a velocity
\begin{equation}
v^{\mu}=\left[-\frac{1}{g_{tt}},-\sqrt{-\left(\frac{1}{g_{tt}g_{rr}}+\frac{1}{g_{rr}}\right)}\;,0,0\right]\;.
\end{equation}
The proper time measured by the observer $dT=-v_{\mu}dx^{\mu}$ is given by
\begin{equation}
dT=dt+g_{rr}\sqrt{-\left(\frac{1}{g_{tt}g_{rr}}+\frac{1}{g_{rr}}\right)}dr\;.
\end{equation}
The metric in the $(T,r,\theta,\phi)$ coordinates is given by
\begin{equation}
ds^2=g_{tt}dT^2+2\sqrt{-g_{tt}g_{rr}(1+g_{tt})}dTdr-g_{tt}g_{rr}dr^2+r^2d\Omega^2\;.
\end{equation}
Since this transformation does not change the normal the requirement for the surface to be null is still $g^{rr}=0$. However, as with the
Eddington-Finkelstein coordinates, this coordinate system is still singular unless $g_{tt}g_{rr}$ is finite and nonzero on the surface.
To obtain a nonsingular coordinate system define a new radial variable by
\begin{equation}
\bar{r}=\int\sqrt{-g_{tt}g_{rr}}dr
\end{equation}
and the metric in this new coordinate system is given by
\begin{equation}
ds^2=g_{tt}dT^2+2\sqrt{1+g_{tt}}\;dTd\bar{r}+d\bar{r}^2+r(\bar{r})^2d\Omega^2\;.
\end{equation}
The condition for the normal $n_{\mu}=(0,1,0,0)$ to be null is $g_{tt}=0$, as expected.

Another approach that can be used to determine the nature of a hypersurface is to examine the induced metric on the
surface. Let the surface have intrinsic coordinates $y^a$ and let the surface be defined by $x^{\mu}=x^{\mu}(y^a)$.
The induced metric $h_{ab}$ is given by
\begin{equation}
h_{ab}=g_{\mu\nu}\frac{\partial x^{\mu}}{\partial y^a}\frac{\partial x^{\nu}}{\partial y^b}\; .
\end{equation}
The induced metric is independent of the space-time coordinates and so the nature of the surface can be deduced in any
coordinate system. All of the coordinate systems used above give
\begin{equation}
h_{ab}=
\left(
  \begin{array}{ccc}
    g_{tt} & 0 & 0 \\
    0 & r_0^2 & 0 \\
    0 & 0 & r_0^2\sin^2\theta \\
  \end{array}
\right)
\end{equation}
for the induced metric on the surface $r=r_0$. Thus, the surface will be null iff $g_{tt}=0$. If $g_{tt}<0$ the
surface will have a Lorentzian signiture and if $g_{tt}>0$ the surface will have a Riemannian signature.

One can also examine the null surface to see if it is singular.
The Kretschmann scalar is given by \cite{Br1}
\begin{equation}
I=R^{\mu\nu\alpha\beta}R_{\mu\nu\alpha\beta}=K_1^2+2K_2^2+2K_3^2+K_4^2
\end{equation}
where
\begin{equation}
K_1=\frac{1}{2g_{rr}}\left[2\frac{g_{tt}^{''}}{g_{tt}}-\left(\frac{g_{tt}^{'}}{g_{tt}}\right)^2-\frac{g_{tt}^{'}g_{rr}^{'}}
{g_{tt}g_{rr}}\right]\; ,
\end{equation}
\begin{equation}
K_2=\frac{g_{tt}^{'}}{g_{tt}g_{rr}r}\; ,
\end{equation}
\begin{equation}
K_3=\frac{g_{rr}^{'}}{g_{rr}^2r}
\end{equation}
and
\begin{equation}
K_4=2\left[\frac{g_{rr}-1}{g_{rr}r^2}\right]\;.
\end{equation}
For a surface to be nonsingular the Kretschmann scalar must be finite on the surface.
Let the surface be at $r=r_0$ and take
\begin{equation}
g_{tt}\sim -A(r-r_0)^n
\label{gtt}
\end{equation}
for $r$ near $r_0$,where $n,A$ and $\alpha$ are constants and $n>0$. From $K_2$ we see that
\begin{equation}
g_{rr}\sim B(r-r_0)^{-m}\;,
\label{grr}
\end{equation}
were $m$ and $B$ are constants and $m\geq 1$. Thus, $g_{rr}$ must diverge on a nonsingular null
surface. Note that $g_{rr}$ can diverge and $g_{tt}$ can be finite and nonzero on a nonsingular
surface. Thus, $g^{rr}=0$ is a necessary but insufficient condition for a surface to be null and
nonsingular.
The only other constraint comes from
$K_1$ and is given by $m\geq 2$ if $m+n\neq 2$. If $m+n=2$ the constraint is weaker.

It is also possible that the Riemann tensor could have a delta function singularity on the surface
while the Kretschmann scalar remains finite as the surface is approached. The delta function in the
Riemann tensor is related to the jump in the derivatives of the metric across the surface \cite{Po1,Ba1}.
Thus, if the derivatives of the metric are continuous across the surface the Riemann tensor will not
contain a delta function singularity on the surface.

Earlier in this section I defined new coordinate systems based on null and time-like geodesics.
It turns out that in some space-times geodesics do not reach the null surface at a finite affine parameter.
To examine this issue consider the time-like component of the geodesic equation, in the
metric (\ref{spherical})
\begin{equation}
\frac{d}{d\tau}\left(g_{tt}\dot{t}\right)=0\;\;\;\;\;\;\;\Rightarrow\;\;\;\;\;\; g_{tt}\dot{t}=-E\;.
\label{21}
\end{equation}
where $\dot{t}=dt/d\tau$, $E>0$ is a constant and I have also used $\tau$ to denote an affine
parameter along both time-like and null geodesics.
The line element for radial motion can be written as
\begin{equation}
g_{tt}\dot{t}^2+g_{rr}\dot{r}^2=\kappa\;.
\label{22}
\end{equation}
where $\kappa=-1$ for time-like geodesics and $\kappa=0$ for null geodesics. Combining equations (\ref{21})
and (\ref{22}) gives
\begin{equation}
g_{tt}g_{rr}\dot{r}^2=\kappa g_{tt}-E^2\;.
\end{equation}
For a surface of infinite redshift with $g_{tt}$ given by (\ref{gtt}) and $g_{rr}$ given by (\ref{grr}) this
equation becomes
\begin{equation}
\dot{r}^2\sim\frac{\kappa}{B}(r-r_0)^m+\frac{E^2}{AB}(r-r_0)^{m-n}\;.
\end{equation}
The last term on the right hand side dominates when $r\simeq r_0$ for $n>0$. The solution to this equation is
given by
\begin{equation}
r(\tau)-r_0=\left[\frac{\pm E(n-m+2)}{2\sqrt{AB}}(\tau+\alpha)\right]^{\frac{2}{n-m+2}}\;\;\;\;\;\; n-m+2\neq 0\;,
\end{equation}
and
\begin{equation}
r(\tau)-r_0=\beta exp\left(-\frac{E\tau}{\sqrt{AB}}\right)\;\;\;\;\;\;\;\;\; n-m+2=0\;,
\end{equation}
where $\alpha$ and $\beta$ are integration constants. Thus, if $n-m+2>0$ both time-like
and null geodesics will reach the null surface at a finite affine parameter, otherwise they will not.

It is interesting to note that this is the same condition that ensures that the null surface is at a finite
value of $\bar{r}=\int\sqrt{-g_{tt}g_{rr}}dr$. Thus, if $n-m+2>0$ the null surface will be at a finite distance from outside points on a constant
time slice in the $(T,\bar{r},\theta,\phi)$ coordinates. It is also interesting to note that the distance to the null surface depends on the
time slicing. For example the distance, on a constant time slice, to the outer horizon in the extremal Reissner-Nordstr{\o}m space-time is infinite
in the $(t,r,\theta,\phi)$
coordinate system but is finite in the $(T,r,\theta,\phi)$ coordinate system.

It is difficult to find null surfaces in a general space-time. However, in a static space-time
with a metric given in (\ref{metric}) the induced metric on a surface $f(\vec{x})=$constant with intrinsic coordinates
$(t,y^1,y^2)$ is given by
\begin{equation}
h_{ab}=\left(
  \begin{array}{cc}
    g_{tt} & 0 \\
    0 & \sigma_{AB} \\
  \end{array}
\right)
\end{equation}
where
\begin{equation}
\sigma_{AB}=\frac{\partial x^{\mu}}{\partial y^A}\frac{\partial x^{\nu}}{\partial y^B}\;g_{\mu\nu}\;\;\;\;\;\;\;\;\; (A,B=1,2)
\end{equation}
is the two dimension ``spatial" part of the induced metric. If the determinant of $\sigma_{AB}$ does not vanish then null surfaces
correspond to $g_{tt}=0$.

\section{The Kerr Space-time}
The Kerr metric is given by
\begin{eqnarray}
ds^2=-\left(\frac{\Delta-a^2\sin^2\theta}{\rho^2}\right)dt^2+\left[\frac{(\Delta+2Mr)^2-\Delta a^2\sin^2\theta}{\rho^2}\right]\sin^2\theta d\phi^2\\
-4\frac{aMr}{\rho^2}\sin^2\theta dtd\phi
+\frac{\rho^2}{\Delta}dr^2+\rho^2d\theta^2\; ,
\end{eqnarray}
where
\begin{equation}
\Delta=r^2-2Mr+a^2\;
\end{equation}
and
\begin{equation}
\rho^2=r^2+a^2\cos^2\theta\;.
\end{equation}
One approach that has been used to find null surfaces involves looking for surfaces of constant $f(r,\theta)$ whose norm is null (see for example \cite{La1,Ad1}).
The only solutions are surfaces of constant $r$ with $g^{rr}=\Delta/\rho^2=0$. This gives the standard result that the null surfaces are at
$\Delta=0$. However, the coordinate system used is singular on these surfaces. To check this result consider the induced metric on a surface
of constant $r$.
\begin{eqnarray}
d\sigma^2=-\left(\frac{\Delta-a^2\sin^2\theta}{\rho^2}\right)dt^2+\left[\frac{(\Delta+2Mr)^2-\Delta a^2\sin^2\theta}{\rho^2}\right]\sin^2\theta d\phi^2
-\frac{4aMr}{\rho^2}\sin^2\theta dtd\phi
+\rho^2d\theta^2
\end{eqnarray}
This surface will be null iff det$(h_{ab})=0$. The determinant of the induced metric is given by
\begin{equation}
det(h_{ab})=-\Delta\rho^2\sin^2\theta\;,
\end{equation}
so that the surface will be null iff $\Delta=0$, which is the standard result.
\section{Conclusion}
In this paper I considered surfaces in a space-time with a Killing vector $\xi^{\alpha}$ that is time-like and hypersurface orthogonal on one side of the surface. It has been argued \cite{Vi1} that the surface is null if $\xi_{\alpha}\xi^{\alpha}\rightarrow 0$ as the surface is approached from the static region.
In a coordinate system adapted to $\xi$ this implies that surfaces with $g_{tt}=0$ are null. I showed that this conclusion is
incorrect and that the condition is
\begin{equation}
g^{ij}g_{tt,i}g_{tt,j}=0
\end{equation}
for a surface of constant $\xi_{\alpha}\xi^{\alpha}$ to have a null normal.
In a static spherically symmetric space-time, with $\partial_rg_{tt}\neq 0$, this condition is $g^{rr}=0$, which is the condition
that would obtain by considering surfaces of constant $r$ in a space-time with a metric given by (\ref{spherical}).

The problem with the condition
$g^{rr}=0$ in a static spherically symmetric space-time is that the coordinate system is singular on the surface.
One can either use a nonsingular coordinate system or examine the induced metric on the surface to determine if it is null.
By using these approaches it was shown that the correct criteria is $g_{tt}=0$. I also showed that null surfaces
correspond to $g_{tt}=0$ in a static space-time if the ``spatial" part of the
induced metric is nonsingular.

For a null surface to be nonsingular the Kretschmann scalar must be finite on the surface. It was shown that, in spherically symmetric space-times, the Kretschmann scalar
will diverge on a null surface unless $g^{rr}\rightarrow 0$ as the surface is approached. It was also shown that $g^{rr}$ can vanish and $g_{tt}$ can be finite and
nonzero on a nonsingular surface. Thus, $g^{rr}=0$ is a necessary but insufficient condition for a surface to be null and nonsingular.
\section*{Acknowledgements}
This research was supported by the  Natural Sciences and Engineering Research
Council of Canada.

\end{document}